\documentclass[a4paper,twocolumn,showpacs,superscriptaddress,floatfix]{revtex4}

\usepackage{graphicx}
\usepackage{epsf}
\usepackage{epsfig}
\usepackage{amssymb,amsmath}
\usepackage[usenames]{color}
\usepackage{amssymb}
\usepackage{times}

\newcommand{\be}{\begin{equation}}
\newcommand{\ee}{\end{equation}}
\newcommand{\bea}{\begin{eqnarray}}
\newcommand{\eea}{\end{eqnarray}}
\newcommand{\nn}{\nonumber}

\font\tenscr=rsfs10 scaled1100
\font\sevenscr=rsfs7 
\font\fivescr=rsfs5 
\skewchar\tenscr='177
\skewchar\sevenscr='177
\skewchar\fivescr='177
\newfam\scrfam
\textfont\scrfam=\tenscr
\scriptfont\scrfam=\sevenscr
\scriptscriptfont\scrfam=\fivescr

\begin{document}

\title{Black hole Area-Angular momentum-Charge inequality in dynamical non-vacuum spacetimes
}

\author{Mar\'{\i}a E. Gabach Cl\'ement}
\affiliation{
Max-Planck-Institut f{\"u}r Gravitationsphysik, Albert Einstein
Institut, Am M\"uhlenberg 1 D-14476 Potsdam Germany 
}

\author{Jos\'e Luis Jaramillo}
\affiliation{
Max-Planck-Institut f{\"u}r Gravitationsphysik, Albert Einstein
Institut, Am M\"uhlenberg 1 D-14476 Potsdam Germany 
}
\affiliation{
Laboratoire Univers et Th\'eories (LUTH), Observatoire de Paris, CNRS, 
Universit\'e Paris Diderot, 92190 Meudon, France
}

\begin{abstract}
We show that the area-angular momentum-charge inequality 
$(A/(4\pi))^2\geq (2J)^2 + (Q_{\mathrm{E}}^2 + Q_{\mathrm{M}}^2)^2$
holds for
apparent horizons of electrically and 
magnetically
charged rotating black holes in generic dynamical and non-vacuum
spacetimes. More specifically, this quasi-local inequality applies to  
axially symmetric closed outermost stably marginally (outer) trapped surfaces,
embedded in non-necessarily axisymmetric black hole spacetimes with 
non-negative cosmological constant and  matter content satisfying 
the dominant energy condition. 
\end{abstract}


\maketitle

\emph{Introduction.~} Isolated stationary black holes in Einstein-Maxwell
theory are completely characterized by their mass $M$, angular momentum $J$ and 
electric and magnetic charges, $Q_{\mathrm{E}}$ and $Q_{\mathrm{M}}$. 
This {\em no hair} property is endorsed by the black hole uniqueness theorems 
leading to Kerr-Newman spacetimes. In these black hole solutions 
a mass-angular momentum-charge inequality enforces a lower bound for $M$.
Such a constraint among $M$, $J$, $Q_{\mathrm{E}}$ and 
$Q_{\mathrm{M}}$ is however lost in the extended Kerr-Newman family,
including singular solutions without a horizon. In this sense, 
the mass-angular momentum-charge inequality follows when
the physical principle of (weak) cosmic censorship, namely the absence of {\em naked
singularities}, is advocated. 
Weak cosmic censorship conjecture provides a {\em dynamical principle} 
aiming at preserving predictability 
and playing a crucial role in our understanding of classical gravitational collapse.
This picture 
motivates the study of extensions of the total mass-angular momentum-charge
inequality to dynamical contexts, something accomplished in 
\cite{Chrusciel:2009ki}
for vacuum axially symmetric 
spacetimes. In more generic scenarios, in 
particular incorporating matter, it is natural to consider a quasi-local
version of the inequality not involving global spacetime quantities (see
\cite{Dain11} for a review).
An appropriate starting point is the area-angular momentum-charge inequality
$(A/(4\pi))^2\geq (2J)^2 + (Q_{\mathrm{E}}^2 + Q_{\mathrm{M}}^2)^2$ also holding 
in the stationary vacuum case. This inequality (for $Q_{\mathrm{M}}=0$)
has been proved to hold for stationary  axisymmetric spacetimes with matter 
in \cite{AnsPfi07,HenAnsCed08,Hennig:2008zy,Ansorg:2010ru}, 
although requiring electro-vacuum in a neighborhood of the horizon.
Regarding the dynamical 
case~\cite{Dain:2010qr,Acena:2010ws,GabachClement:2011kz,Dain:2011pi}, 
a proof has been presented for the non-vacuum uncharged case
\cite{Jaramillo:2011pg} and the area-charge inequality \cite{Dain:2011kb}
(in absence of any symmetries). Here we extend the full 
area-angular momentum-charge inequality, in particular incorporating the magnetic charge,
to generic non-axisymmetric dynamical non-vacuum black hole spacetimes 
(axial symmetry only required on the horizon). 
This completes the discussion
of this inequality in the Einstein-Maxwell context.

\vspace{0.3cm}
\emph{The result.~} The area-angular momentum-charge inequality applies to
horizon sections satisfying a stability condition.
Following the approach in \cite{Jaramillo:2011pg}, we consider a
closed marginally outer trapped surface ${\cal S}$ satisfying a (spacetime)
{\em stably outermost} condition in the sense of \cite{AndMarSim05,AndMarSim08}
(see Definition 1 below for details).
Then the following result holds:

\vspace{0.2cm}
{\bf Theorem 1.~} {\em Given an axisymmetric closed marginally trapped
surface ${\cal S}$ satisfying the (axisymmetry-compatible)
spacetime stably outermost condition, in a spacetime 
with non-negative cosmological constant and matter
content fulfilling the dominant energy condition, 
it holds the inequality
\bea
\label{e:inequality}
\left(A/(4\pi)\right)^2\geq (2 J)^2 + (Q_{\mathrm{E}}^2 + Q_{\mathrm{M}}^2)^2
\eea
where $A$ is the area of ${\cal S}$ and $J$, $Q_{\mathrm{E}}$ and $Q_{\mathrm{M}}$ are, respectively,
the total (gravitational and electromagnetic) angular momentum, the electric
and the magnetic charges associated with ${\cal S}$. 
}

\vspace{0.3cm}
This quasi-local result holds in fully dynamical spacetimes without bulk symmetries
and with arbitrary (non-exotic) matter possibly crossing the horizon.
In particular, it extends to generic scenarios the inequality 
proved in \cite{Hennig:2008zy,Ansorg:2010ru} for Killing horizons in 
stationary axisymmetric spacetimes, with electrovacuum
around the black hole (matter can surround but not cross the horizon).
Axisymmetry is required only on ${\cal S}$, so that a canonical notion of angular
momentum $J$ can be employed. 
The stably outermost and dominant energy conditions imply,
for some non-vanishing $J$, $Q_{\mathrm{E}}$ or $Q_{\mathrm{M}}$
and in our four-dimensional spacetime context, the spherical topology
of the surface ${\cal S}$. For Killing horizons \cite{AnsPfi07,Hennig:2008zy,Ansorg:2010ru}
 a rigidity result holds, namely
equality in (\ref{e:inequality}) implies the degeneracy of the
Killing horizon (vanishing of the {\em surface gravity}), providing a 
characterization of extremality. In the present dynamical setting, 
with no spacetime stationary Killing field, rigidity statements 
involve rather the characterization of the induced metric on 
${\cal S}$ as an {\em extremal throat} (i.e. with the geometry of a horizon section 
in the extremal Kerr-Newman family) and as 
a section of an instantaneous (non-expanding) isolated horizon \cite{AshKri04}.
We postpone the discussion of the rigidity part
of the result to \cite{DaiGabJar11}, where full details of the proof of inequality
(\ref{e:inequality}) [required to make the rigidity statement precise]
are presented.

\vspace{0.3cm}
\emph{Main geometric elements.~} 
The proof of (\ref{e:inequality}) proceeds by, first, casting the stably outermost
condition for marginally outer trapped surfaces as a geometric inequality
leading to an action functional ${\cal M}$ on ${\cal S}$
and, second, by solving the associated variational problem.
Following \cite{Jaramillo:2011pg},
we start by introducing the general geometric elements
and by formulating the geometric inequality following from the stability
of ${\cal S}$. 

Let $(M, g_{ab})$ be a $4$-dimensional spacetime
with Levi-Civita connection $\nabla_a$, satisfying the dominant energy 
condition and with non-negative cosmological constant $\Lambda\geq 0$. 
Let us consider an electromagnetic field with 
strength field (Faraday) tensor $F_{ab}$, so that $F_{ab}=\nabla_a A_b - \nabla_b A_a$
on a local chart (corresponding to a given section of the $U(1)$-fibre-bundle, 
possibly non-trivial to account for magnetic monopoles).  

Let us consider a closed orientable 2-surface ${\cal S}$ embedded in $(M, g_{ab})$.
Regarding its intrinsic geometry, let us denote the
induced metric as $q_{ab}$  with  connection $D_a$, 
Ricci scalar as ${}^2\!R$,  volume element $\epsilon_{ab}$ and 
area measure $dS$. Regarding its extrinsic geometry, 
we first consider normal (respectively, outgoing and ingoing)
null vectors $\ell^a$ and $k^a$ normalized as $\ell^a k_a = -1$. 
This fixes $\ell^a$ and $k^a$ up to
(boost) rescaling factor. The  extrinsic curvature elements 
needed in our analysis are the expansion $\theta^{(\ell)}$, the shear
$\sigma^{(\ell)}_{ab}$ and the normal fundamental form $\Omega_a^{(\ell)}$  
associated with the outgoing null normal $\ell^a$
\bea
\label{e:expansion_shear}
\theta^{(\ell)}&=&q^{ab}\nabla_a\ell_b \ \ , \ \  
\sigma^{(\ell)}_{ab}=  {q^c}_a {q^d}_b \nabla_c \ell_d - \frac{1}{2}\theta^{(\ell)}q_{ab} \nn \\
\Omega^{(\ell)}_a &=& -k^c {q^d}_a \nabla_d \ell_c \ .
\eea
We require the geometry of ${\cal S}$ to be axisymmetric
with axial Killing vector $\eta^a$ on ${\cal S}$. That is, ${\cal L}_\eta q_{ab}=0$
and $\eta^a$ has closed integral curves, vanishes exactly 
at two points on ${\cal S}$ and is normalized so that its integral curves 
have an affine length of $2\pi$. Besides, we demand ${\cal L}_\eta \Omega_a^{(\ell)}=
{\cal L}_\eta A_a=0$ and adopt a tetrad  $(\xi^a, \eta^a, \ell^a, k^a)$ 
on ${\cal S}$
adapted to axisymmetry, namely ${\cal L}_\eta \ell^a = {\cal L}_\eta k^a = 0$
with $\xi^a$ a unit vector tangent to ${\cal S}$ satisfying
$\xi^a\eta_a=\xi^a\ell_a=\xi^ak_a=0$, $\xi^a\xi_a=1$.
We can then write 
$q_{ab}=\frac{1}{\eta}\eta_a\eta_b + \xi_a\xi_b$
(with $\eta=\eta^a\eta_a$) and $\Omega^{(\ell)}_a = \Omega^{(\eta)}_a + \Omega^{(\xi)}_a$
(with  $\Omega^{(\eta)}_a= \eta^b\Omega^{(\ell)}_b \eta_a/\eta$
and $\Omega^{(\xi)}_a= \xi^b\Omega^{(\ell)}_b \xi_a$).

We introduce now the expressions for
$J$, $Q_{\mathrm{E}}$ and $Q_{\mathrm{M}}$. First, the electric 
and magnetic field components normal to ${\cal S}$ are
\bea
\label{e:E_B_normal}
E_\perp = F_{ab}\ell^ak^b \ \ \ , \ \ \ B_\perp = {}^*\!F_{ab}\ell^ak^b \ ,
\eea
where ${}^*\!F_{ab}$ is the Hodge dual of $F_{ab}$.
The above-required axisymmetry allows the introduction of the 
following canonical notion of angular momentum 
on ${\cal S}$ \cite{Carte10,Simon:1984qb,Ashtekar:2001is, Dain11}
\bea
\label{e:angular_momentum}
J = J_{_{\mathrm{K}}} + J_{_{\mathrm{EM}}} =\frac{1}{8\pi}\int_{\cal S} \Omega_a^{(\ell)} \eta^a dS 
+ \frac{1}{4\pi}\int_S (A_a \eta^a) E_\perp dS \ ,
\eea
where $J_{_{\mathrm{K}}}$ and $J_{_{\mathrm{EM}}}$ correspond, respectively,
to (Komar) gravitational and  electromagnetic contributions to the total $J$.
Electric and magnetic charges can be expressed as (e.g. \cite{AshFaiKri00,Booth:2007wu})
\bea
\label{e:charges}
Q_{\mathrm{E}}= \frac{1}{4\pi}\int_S E_\perp dS \ , \  
Q_{\mathrm{M}}=\frac{1}{4\pi}\int_S B_\perp dS \ .
\eea

We characterize now ${\cal S}$ as a stable section of a
(quasi-local) black hole horizon. 
First, we require ${\cal S}$ to be a 
marginally outer trapped surface, that is $\theta^{(\ell)}=0$.
Second, we demand ${\cal S}$ to be stably outermost as
introduced in  \cite{AndMarSim05,AndMarSim08}
(see also \cite{Hay94,Racz:2008tf}). More specifically
we require ${\cal S}$ to be
(axisymmetry-compatible) {\em spacetime stably outermost} 
\cite{Jaramillo:2011pg,Dain:2011kb}:

\vspace{0.2cm}
{\bf Definition 1.~} {\em A closed marginally trapped surface ${\cal S}$ is referred
to as {\em spacetime stably outermost} if there exists 
an outgoing ($-k^a$-oriented) vector $X^a= \gamma \ell^a - 
\psi k^a$, with $\gamma\geq0$ and $\psi>0$, with respect to which 
${\cal S}$ is stably outermost: $\delta_X \theta^{(\ell)} \geq 0$.
If, in addition, $X^a$ (i.e. $\gamma$, $\psi$)
and $\Omega^{(\ell)}_a$ are axisymmetric,
we will refer to $\delta_X \theta^{(\ell)}\geq 0$ as an (axisymmetry-compatible) 
spacetime stably outermost condition.
}
\vspace{0.2cm}

Here, the operator $\delta_X$ is the variation operator on the surface ${\cal S}$
along the vector $X^a$ discussed in \cite{AndMarSim05,AndMarSim08} 
(see also \cite{BooFai07,Cao:2010vj}).
We formulate now the following lemma:
\vspace{0.2cm}

{\bf Lemma 1.~}
{\em Let ${\cal S}$ be a closed marginally trapped surface ${\cal S}$ satisfying the 
(axisymmetry-compatible) spacetime stably outermost condition. 
Then, for all axisymmetric $\alpha$ on ${\cal S}$
\bea
\label{e:geom_inequality}
\int_{\cal S} \left[|D \alpha|^2 + \frac{1}{2} \alpha^2 
\; {}^2\!R \right] dS 
\geq \int_{\cal S} \alpha^2 \left[|\Omega^{(\eta)}|^2 
+ (E_\perp^2 + B_\perp^2)\right]dS, 
\eea
with $|D \alpha|^2 = D_a\alpha D^a\alpha$
and $|\Omega^{(\eta)}|^2=\Omega^{(\eta)}_a  {\Omega^{(\eta)}}^a$.
}

\vspace{0.2cm}
The proof is a direct application of Lemma 1 in \cite{Jaramillo:2011pg}.
Given the vector $X^a= \gamma \ell^a - \psi k^a$,
for all $\alpha$ on ${\cal S}$ it holds \cite{Jaramillo:2011pg}
\bea
\label{e:inequality_alpha}
&&\int_{\cal S} \left[D_a\alpha D^a\alpha + \frac{1}{2} \alpha^2 
\; {}^2\!R \right] dS \geq
 \\ 
&& \int_{\cal S} \left[ \!\alpha^2 \Omega^{(\eta)}_a  {\Omega^{(\eta)}}^a +
\alpha \beta \sigma^{(\ell)}_{ab} {\sigma^{(\ell)}}^{ab} 
+ G_{ab}\alpha\ell^a (\alpha k^b + \beta\ell^b) \right] dS \nn \ ,
\eea
with $\beta=\alpha\gamma/\psi$. First, since $\alpha\beta \geq 0$,
the positive-definite quadratic term in the shear can be neglected.
Second, we insert Einstein equation $G_{ab} + \Lambda g_{ab} = 8\pi(T_{ab}^{\mathrm{EM}}
+T_{ab}^{\mathrm{M}})$,
with $T_{ab}^{\mathrm{EM}}$ and $T_{ab}^{\mathrm{M}}$ the electromagnetic and matter stress-energy 
tensors. 
In particular, 
$T_{ab}^{\mathrm{EM}}=\frac{1}{4\pi}\left(F_{ac}F_b{}^{c}-\frac{1}{4}g_{ab} F_{cd} F^{cd}\right)$.
From the dominant energy condition on $T_{ab}^{\mathrm{M}}$, $\Lambda\geq 0$ and 
the null energy condition applying for $T_{ab}^{\mathrm{EM}}$,
the Einstein tensor term in inequality (\ref{e:inequality_alpha})
is bounded by below by $\alpha^2 8\pi T_{ab}^{\mathrm{EM}}\ell^ak^b$. Making use of 
(see e.g. \cite{Booth:2007wu,Dain:2011kb})
\bea
\label{e:Tlk}
T^{\mathrm{EM}}_{ab}\ell^ak^b = \frac{1}{8\pi} 
    \left[\left(\ell^a k^b F_{ab}\right)^2 
    +\left(\ell^a k^b {}^*\!F_{ab}\right)^2\right] \ ,
\eea
inequality (\ref{e:geom_inequality}) follows by identifying $E_\perp$ and $B_\perp$ in
(\ref{e:E_B_normal}). As a final remark, note that taking $\alpha=\mathrm{const}$
in  (\ref{e:geom_inequality}),
a non-vanishing angular momentum or charge suffices to conclude
the sphericity of ${\cal S}$ by applying the Gauss-Bonnet theorem.

\vspace{0.3cm}
\emph{The action functional and sketch of the proof.~} 
The proof of inequality (\ref{e:inequality}) proceeds by solving a constrained
variational problem on ${\cal S}$, in which 
$J$, $Q_{\mathrm{E}}$ and $Q_{\mathrm{M}}$ must be kept constant under otherwise arbitrary variations.
We construct the corresponding action functional ${\cal M}$, 
by evaluating the  geometric expression (\ref{e:geom_inequality})
in a specific coordinate system on ${\cal S}$. 

First, on an axisymmetric sphere ${\cal S}$, a coordinate system can always
be chosen such that 
\bea
\label{e:q_ab}
ds^2=q_{ab}dx^a dx^b = e^\sigma \left(e^{2q} d\theta^2 + 
\mathrm{sin}^2\theta d\varphi^2 \right) \ ,
\eea
with axisymmetric $\sigma$ and $q$ satisfying $\sigma+q=c=\mathrm{constant}$. 
Then $\eta^a=(\partial_\varphi)^a$,  $\eta = e^\sigma \mathrm{sin}^2\theta$ 
and $dS=e^c dS_0$, 
with  $dS_0= \mathrm{sin}\theta d\theta d\varphi$. In particular, $A=4\pi e^c$.
Second, $\Omega^{(\ell)}_a$ expresses uniquely
on a 2-sphere as
$\Omega^{(\ell)}_a = \epsilon_{ab} D^b \tilde{\omega} + D_a\lambda$. Since  $\Omega^{(\ell)}_a$
is axisymmetric, $\Omega^{(\eta)}_a=\epsilon_{ab} D^b \tilde{\omega}$  \cite{Jaramillo:2011pg}, 
and we can write
\bea
\label{e:Omega_eta}
\Omega^{(\eta)}_a = \frac{1}{2\eta}\epsilon_{ab} D^b \bar{\omega} \ ,
\eea
by introducing the potential $\bar{\omega}$, as 
$d\bar{\omega}/d\theta= (2\eta) d\tilde{\omega}/d\theta$, that  
satisfies $J_{_{\mathrm{K}}}=[\bar{\omega}(\pi)- \bar{\omega}(0)]/8$ 
(cf. \cite{Jaramillo:2011pg}).
Third, from 
$\ell^a k^b {}^*\!F_{ab}=\frac{1}{2}F_{ab}\epsilon^{ab}$ \cite{Dain:2011kb} and
the axisymmetry of $A_a$
\bea
\label{e:B_normal}
B_\perp = \frac{1}{e^c \sin \theta} \frac{dA_\varphi}{d\theta} \ .
\eea
Finally, following \cite{Dain:2011pi,Jaramillo:2011pg} we choose
$\alpha = e^{c-\sigma/2}$. Inserting it together with 
(\ref{e:q_ab}), (\ref{e:Omega_eta}), (\ref{e:B_normal}) 
into inequality (\ref{e:geom_inequality}), we get
\bea
\label{e:ineqM}
8(c+1)\geq {\cal M}[\sigma, \bar{\omega}, E_\perp, A_\varphi] \ ,
\eea
where ${\cal M}[\sigma, \bar{\omega}, E_\perp, A_\varphi]$ is the action functional 
\bea
\label{e:functional_v1}
{\cal M}[\sigma, \bar{\omega}, E_\perp, A_\varphi] =
\frac{1}{2\pi}\int_{\cal S}
\left[4\sigma+
\left(\frac{d\sigma}{d\theta}\right)^2+\frac{1}{\eta^{2}}
\left(\frac{d\bar{\omega}}{d\theta}\right)^{2} \right. \\
\!\!\!\left. + 4e^{2c-\sigma}E_\perp^2 + 4e^{-\sigma} 
\!\!\left(\frac{1}{\mathrm{sin}\theta}
\!\!\frac{dA_\varphi}{d\theta}\right)^2 \right]dS_0. \nn
\eea
Inequality (\ref{e:inequality}) follows by solving the
variational problem defined by ${\cal M}[\sigma, \bar{\omega}, E_\perp, A_\varphi]$.
In its form (\ref{e:functional_v1}), enforcing
the constraints on $J$, $Q_{\mathrm{E}}$ and $Q_{\mathrm{M}}$ is not straightforward. 
This is addressed by
introducing new potentials $\omega$, $\chi$ and $\psi$ on ${\cal S}$
\bea
\label{e:omega_chi_psi}
\frac{d\psi}{d\theta}&=&E_\perp e^c\sin\theta \ \ \ \ , \ \ \ \ \chi=A_\varphi \ , \nn \\
\frac{d\omega}{d\theta}&=& 2\eta\frac{d\tilde\omega}{d\theta}
+2\chi\frac{d\psi}{d\theta}-2\psi\frac{d\chi}{d\theta} \ ,
\eea
with the crucial property that $J$, $Q_{\mathrm{E}}$ and $Q_{\mathrm{M}}$ are written as
\bea
\label{e:J_QE_QM_omega_chi_psi}
J=\frac{\omega(\pi)-\omega(0)}{8}, Q_{\mathrm{E}}=\frac{\psi(\pi)-\psi(0)}{2},
Q_{\mathrm{B}}=\frac{\chi(\pi)-\chi(0)}{2}.
\eea
Physical parameters in inequality (\ref{e:inequality})
can then be kept constant by fixing  
$\omega$, $\chi$ and $\psi$ on the axis as a boundary condition in the variational
problem (note that $\bar\omega$ in (\ref{e:Omega_eta}) is an appropriate 
potential to control the Komar $J_{_{\mathrm{K}}}$, but not for the total $J$).
In terms of $\sigma$,  $\omega$, $\chi$ and $\psi$ the action functional reads
\bea
\label{e:functional_v2}
&&\!\!\! {\cal M}[\sigma, \omega, \psi, \chi]=
\frac{1}{2\pi}\int_{\cal S}\left[ 4\sigma+|D\sigma|^2 \right. \\
&&\left. + \frac{|D\omega-2\chi D\psi+2\psi D\chi|^2}{\eta^2}
+\frac{4}{\eta}(|D\psi|^2+|D\chi|^2) \right]dS_0 \nn \ ,
\eea
where ${\cal M}$ is formally promoted beyond axisymmetry.
The proof of (\ref{e:inequality}) proceeds in two steps
(see details in \cite{DaiGabJar11}). 
First
\bea
\label{e:AgeqeM}
A\geq 4\pi e^{\frac{{\cal M}-8}{8}} \ ,
\eea
follows directly from (\ref{e:ineqM}) and $A=4\pi e^c= 4\pi e^{\sigma(0)}$. Second,
by solving the variational problem defined by the action functional
(\ref{e:functional_v2}) with values of $\omega, \psi, \chi$ fixed on the
axis and determined from relations (\ref{e:J_QE_QM_omega_chi_psi}), it is shown 
\bea
\label{e:MgeqM0}
{\cal M}\geq {\cal M}_0 = 8\ln\sqrt{(2J)^2 + (Q_{\mathrm{E}}^2+ Q_{\mathrm{M}}^2)^2} + 8 \ , 
\eea
where ${\cal M}_0$ corresponds to the evaluation of
${\cal M}$ on an extremal solution in the (magnetic) Kerr-Newman family
with given $J$, $Q_{\mathrm{E}}$ and $Q_{\mathrm{M}}$. Inequality (\ref{e:inequality})
follows from the combination of inequalities (\ref{e:AgeqeM}) and (\ref{e:MgeqM0}).
Full intermediate details of the proof, in particular addressing 
the resolution of the variational problem along the lines in \cite{Acena:2010ws}
will be presented in \cite{DaiGabJar11}.

\vspace{0.3cm}
\emph{Explicit proof of the vanishing magnetic charge case.~}
Complementary to the discussion above of the 
elements in the proof of the full inequality (\ref{e:inequality}),
we present a straightforward explicit proof of the case 
$Q_{\mathrm{M}}=0$ by matching the reasoning
in \cite{Hennig:2008zy}. The result in \cite{Hennig:2008zy} states 
that a {\em subextremal} stationary
black hole, in the sense that trapped surfaces exist in the interior vicinity of the
event horizon \footnote{Horizon sections are {\em strictly stably outermost} 
with respect to outgoing directions (namely, 
{\em outer trapping horizons} \cite{Hay94,Booth:2007wu}).}, 
satisfies the strict inequality (\ref{e:inequality}).
Namely, 
\bea
\label{e:HCA}
\hbox{{\em horizon subextremal condition}} \Rightarrow
p_J^2 + p_Q^2<1 \ .
\eea
where $p_J = \frac{8\pi J}{A}$ and $p_Q = \frac{4\pi Q_{\mathrm{E}}^2}{A}$.
This implication (actually, its logical counter-reciprocal) is cast in \cite{Hennig:2008zy} 
as a variational problem on a Killing horizon section.
The action functional in \cite{Hennig:2008zy} is constructed by combining the 
{\em horizon subextremal condition} in (\ref{e:HCA}) with the condition $p_J^2 + p_Q^2<1$.
The key remark here is to show that such variational problem, 
defined solely on a sphere ${\cal S}$,
has full applicability in the generic dynamical case beyond the
original stationary and spacetime axisymmetric setting of \cite{Hennig:2008zy}.
More specifically, we show that our expressions for $p_J$, $p_Q$
and the stably outermost condition (\ref{e:ineqM}),
valid in the generic dynamical non-vacuum case,
match exactly the expressions in \cite{Hennig:2008zy} for the elements
in  (\ref{e:HCA}). Therefore, the proof in \cite{Hennig:2008zy} 
extends {\em exactly} to the generic case.

From the comparison between the 4-dimensional stationary axisymmetric line element in 
\cite{Hennig:2008zy} with our line element (\ref{e:q_ab}) on ${\cal S}$ and
between the respective integrands of the Komar angular momentum,
we introduce new fields $U$ and $V$ from $\sigma$ and $\bar\omega$
\bea
\label{e:UV}
\hat u&=&e^\sigma \ , \ \hat u_N=e^c \ , \  U=\frac{1}{2}\ln\left(\frac{\hat u}{\hat u_N}\right)
\ , \nn \\ 
V &=& \frac{e^\sigma \sin\theta}{2\eta^2}\frac{d\bar\omega}{d\theta} \ .
\eea
Regarding the electromagnetic potentials, we define $S$ and $T$
\bea
\label{e:ST}
S=-E_\perp e^{c/2} \ \ \ , \ \ \ T=A_\varphi e^{-c/2} \ .
\eea
Inserting these fields in (\ref{e:angular_momentum}) and (\ref{e:charges}) above,
using $A=4\pi e^c$ and changing to variable $x=\cos\theta$ we get
\bea
\label{e:p_J_p_Q}
p_J &=& -\frac{1}{2}\int_{-1}^1 V e^{2U}(1-x^2) dx + \int_{-1}^1 S T dx \nn \\
p_Q &=&  \frac{1}{4}\left(\int_{-1}^1 S dx\right)^2 \ ,
\eea
that coincide exactly with expressions in Eqs. (23) and (24) in \cite{Hennig:2008zy}.
Regarding the stability (subextremal) condition, we insert (\ref{e:UV})
and (\ref{e:ST}) in condition (\ref{e:ineqM}) [with strict inequality].
Using  $\int_{-1}^1 Udx=-\int_{-1}^1 U'xdx$ (following from
 $U(1)=U(-1)=0$, as a regularity condition for $q$ on the axis)
and denoting with a prime the derivative with respect to $x$, we find
\bea
\label{e:subextremal_cond}
1 >\frac{1}{2}\int_{-1}^1(U'^2+V^2)(1-x^2)-2U'x+e^{-2U}(S^2+T'^2)dx.
\eea 
This  matches exactly the {\em horizon subextremal condition} inequality
(28) in \cite{Hennig:2008zy}.
Considering expressions (\ref{e:p_J_p_Q}) and (\ref{e:subextremal_cond}), 
the same variational problem used in the proof of (\ref{e:HCA})
can be defined in the generic case. This is a complete proof 
of inequality (\ref{e:inequality}) with vanishing $Q_{\mathrm{M}}$ in the
strictly stably case.

\vspace{0.3cm}
\emph{Discussion.~}
We have shown that 
$(A/(4\pi))^2\geq (2\pi J)^2 + (Q_{\mathrm{E}}^2 + Q_{\mathrm{M}}^2)^2$
holds for axisymmetric stable marginally trapped surfaces 
in generically dynamical, non-necessarily axisymmetric spacetimes with 
ordinary matter that can be crossing the horizon. More specifically,
we have presented a complete proof of the strictly stable case 
with $Q_{\mathrm{M}}=0$
and provided the key elements for the proof
of the general inequality.
From the perspective of the {\em no hair} property of vacuum stationary 
black holes, the extension of inequality (\ref{e:inequality}) to fully dynamical
non-vacuum situations represents a remarkable result.
Indeed, although parameters $A, J, Q_{\mathrm{E}}$ and $Q_{\mathrm{M}}$
do not longer fully characterize the black hole state and new degrees of freedom
are required to describe the spacetime geometry, the generic incorporation of the
latter is still constrained by inequality (\ref{e:inequality}).
Such a constraint represents a valuable probe into non-linear black hole dynamics.
As a first remark, it gives support to the physical interpretation 
of the Christodoulou mass in  
dynamical settings (cf. discussion in \cite{Dain11}), in particular 
endorsing dynamical horizon~\cite{AshKri04}
thermodynamics~\footnote{In this context, we note that inequalities
(\ref{e:geom_inequality}) and  (\ref{e:inequality_alpha}) can be interpreted as
upper bounds on certain {\em energy fluxes} defined by their right-hand-sides 
(and closely related to dynamical horizon fluxes
\cite{AshKri02,AshKri03}). 
In particular, terms proportional
to $\beta$ in (\ref{e:inequality_alpha}) correspond to gravitational and electromagnetic 
radiative degrees of freedom ($T^{\mathrm{EM}}_{ab}\ell^a\ell^b$ being the flux 
of the Poynting vector).}.
More generally, whereas  inequality (\ref{e:inequality}) follows originally  
in the Kerr-Newman family under the assumption of (weak)
cosmic censorship, 
the present result is purely quasi-local involving no global condition on the spacetime,
namely no asymptotic predictability.
This suggests a link between cosmic censorship and  marginally trapped surface 
stability to be further explored.
In this context, assuming  Penrose inequality 
(with no surface enclosing ${\cal S}$ with area smaller than $A$), 
inequality (\ref{e:inequality}) refines the positive of mass 
theorem in terms of physical quantities: $16\pi M^2\geq A \geq 
\sqrt{(8\pi J)^2 + (4\pi [Q_{\mathrm{E}}^2 + Q_{\mathrm{M}}^2])^2}$. Although for non-axisymmetric
horizons we lack a canonical notion of angular momentum, appropriate quasi-local
prescriptions for $J$ should provide good estimates for a lower bound of $M$. 
Giving closed general expressions seems however
difficult since, in contrast with the area-charge inequality \cite{Dain:2011kb},
incorporating $J$ involves a subtle variational problem (cf.
\cite{Dain11}). In this sense, 
Ref.  \cite{DaiGabJar11} discusses the
close relation between the variational problem (on a 3-slice) 
employed in \cite{Chrusciel:2009ki} for the proof of the spacetime 
mass-angular momentum-charge inequality and
the present action functional ${\cal M}$ 
in (\ref{e:functional_v1}) and (\ref{e:functional_v2}), also 
closely related to (but different from) the functional
used in \cite{Hennig:2008zy}. Regarding the latter, we stress
that electromagnetic potentials $S$ and $T$ in (\ref{e:ST}) 
follow straightforwardly
(with no gauge choices involved) from the  geometric formulation of 
the general stability condition in Lemma 1.
This underlines the intrinsic interest 
of the {\em flux inequality} in Lemma 1 (and, more generally, its complete expression in 
\cite{Jaramillo:2011pg}) for exploring further geometric aspects of stable black hole horizons.


\noindent\emph{Acknowledgments.~} 
We thank S. Dain, M. Reiris and W. Simon for the in-depth discussion of
crucial aspects of this work and for their encouraging support.
We would like  also to thank A. Ace\~na, P. Aguirre and M. Ansorg 
for enlightening discussions.
J.L.J. acknowledges the Spanish MICINN 
(FIS2008-06078-C03-01) and the Junta de Andaluc\'\i a (FQM2288/219).






\end{document}